\documentclass[%
reprint,
superscriptaddress,
amsmath,amssymb,
prl,
]{revtex4-2}

\usepackage{graphicx}
\graphicspath{{figures/}}

\usepackage{dcolumn}
\usepackage{bm}

\usepackage[dvipsnames]{xcolor}

\newcommand{\tu}[1]{^\text{\tiny #1}} 
\newcommand{\tb}[1]{_\text{\tiny #1}} 
\newcommand{\C}{_\text{\tiny c}}

\newcommand{\kR}{k\tb{R}}
\newcommand{\kDh}{k\tb{D}\tu{h}}
\newcommand{\kDg}{k\tb{D}\tu{g}}

\usepackage[normalem]{ulem} 

\usepackage[hidelinks]{hyperref}

\usepackage{xr}
\begin{document}
	
	\preprint{APS/123-QED}
	
	\title{Synergistic chemomechanical dynamics of feedback-controlled microreactors}
    
	
	\author{Sebastian Milster}
	\affiliation{Applied Theoretical Physics - Computational Physics, Physikalisches Institut, Albert-Ludwigs-Universit\"at Freiburg, D-79104 Freiburg, Germany}
		\author{Abeer Darwish}
	\affiliation{Applied Theoretical Physics - Computational Physics, Physikalisches Institut, Albert-Ludwigs-Universit\"at Freiburg, D-79104 Freiburg, Germany}
	\author{Nils G\"oth}
	\affiliation{Applied Theoretical Physics - Computational Physics, Physikalisches Institut, Albert-Ludwigs-Universit\"at Freiburg, D-79104 Freiburg, Germany}
   	\author{Joachim Dzubiella}
	\affiliation{Applied Theoretical Physics - Computational Physics, Physikalisches Institut, Albert-Ludwigs-Universit\"at Freiburg, D-79104 Freiburg, Germany}
    \affiliation{Cluster of Excellence livMatS @ FIT - Freiburg Center for Interactive Materials and Bioinspired Technologies, Albert-Ludwigs-Universit\"at Freiburg, D-79110 Freiburg, Germany}
	
	\date{\today}

\begin{abstract}

The experimental control of synergistic chemomechanical dynamics of catalytically active microgels (microreactors) is a key prerequisite for the design of adaptive and biomimetic materials. Here, we report a minimalistic model of feedback-controlled microreactors based on the coupling between the hysteretic polymer volume phase transition and a volume-controlled permeability for the internal chemical conversion. We categorize regimes of mono- and bistability, excitability, damped oscillations, as well as sustained oscillatory states with tunable amplitude, as indicated by experiments and representable by the FitzHugh-Nagumo dynamics for neurons. We summarize the features of such a `colloidal neuron' in bifurcation diagrams with respect to microgel design parameters, such as permeability and relaxation times,  as a guide for experimental synthesis.
\end{abstract}

\maketitle





%
%
{\it Introduction.---}In all living matter, the switching between states, sustained oscillations, or excitability to external stimuli are integral to biofunctional processes, such as spatiotemporal self-organization, intercellular communication, homeostasis, memory and learning, and collective decision making~\cite{Barkai,homeo,Aizenberg,Goldbeter}. These features typically exploit dissipative biochemical feedback loops across multiple length and time scales, inspiring new dynamic functionalities and pathways in the design of bioinspired and 'smart' synthetic materials~\cite{Walther2020,Ikkala,Duzs, Levine1}. In particular, {\it chemomechanical transduction}, i.e., the transformation of chemical into mechanical energy through oscillations or waves, or {\it vice versa}~\cite{Yashin2012,hydrogelsignalling}, is a key prerequisite for mechanical adaption and signal propagation in smart materials~\cite{Maity2021,Huber2022}. 

Recent research is diving into the challenging realm of {\it synergistic} chemomechanical dynamics at the {\it microscale}, and is increasingly employing synthetic, stimuli-responsive microgels because of their versatile tunability. Catalytically active hydrogels (nano- or microreactors~\cite{Rafa}) are particularly promising for applications since the volume phase transition (VPT), and physicochemical properties, such as diffusion and reaction timescales, are well controllable by rational polymer design~\cite{Stuart,Nanogelreview}.  Among the numerous dynamical behaviors, synergistic oscillations are remarkably fascinating, where in contrast to the simpler `enslaved' microgels (driven by intrinsically oscillating chemical reactions~\cite{Sakai,Yashin2012,Yoshida1999,Crook2002, Straube}), neither the chemical reaction nor the polymer alone exhibit oscillations. Only by chemomechanical coupling, and under the prerequisite of a {\it hysteretic} VPT, sustained oscillations may result~\cite{Baker1996,Zou1999}. 

Synergistic oscillations can in principle be achieved by the negative feedback and self-regulation mediated through the bistable switching of the microgel {\it permeability}, triggered by the rising concentration of products as, e.g., in glucose~\cite{Katchalsky1972, Hahn1973, Siegel1995, Baker1996, Leroux1999, Zou1999, Dhanarajan2002, Bhalla, Kepperbook} or bromate-sulfite driven pH-induced VPTs~\cite{Boissonade2009,Horvath}. However, the few reported clear experimental realizations required either very complex hierarchies~\cite{Aizenberg} or took place only on macroscopic scales~\cite{Dhanarajan2002, Horvath, Bhalla}.  Only recently, indications of chemomechanical oscillators based on permeability changes were reported for spherical, catalytically active microgels on smaller scales~\cite{Narita2013, Bell2021a}. However, for the desired {\it miniaturization} to milli- to microscales the precise control and rational guidance of the experimental process parameters seems essential to achieve robust oscillations.

In this Letter, we present such a theoretical  guiding map for the stationary dynamics of a permeability-controlled chemomechanical microreactor. Remarkably, our model is mathematical coherent with the well-studied FitzHugh-Nagumo (FHN) dynamics for excitable media, such {as chemical oscillators~\cite{Nitzan1974,Boissonade1980,Epstein1984} and} neurons~\cite{Fitzhugh,Scott}. This allows an amenable stability analysis as well as the extrapolation to collective behavior of coupled as well as {\it  noisy} oscillators~\cite{Lindner,Levine2,Neural,chimera,PhysRevLett.130.107202,PhysRevLett.130.107201} for smart material design in future. Our model of a `colloidal neuron' converges all essential ingredients as put forward in previous pioneering works {\cite{Hahn1973,Boissonade2009,Yashin2012,Kepperbook, Katchalsky1972, Hahn1973, Siegel1995, Baker1996, Leroux1999, Zou1999, Dhanarajan2002, Bhalla,Yao2017}} using more elaborate models into a minimalist framework. It thus allows a holistic description of the robust phenomena observed for various chemicals and geometries, while representing the same physics.  In particular, we explicitly consider the microgel relaxation dynamics within a generic bistable Landau-like energy landscape~\cite{Landau} for the hysteretic VPT~\cite{Suzuki,Drozdov} to include a Flory-Rehner like elastic response~\cite{Flory,Boissonade2005,Yashin2012}, an exponential volume-dependent permeability (sieving) function~\cite{Gehrke, Dhana}, as supported by recent simulations~\cite{Q,Matej}, and the (in meaningful limits) adequate first order rate equations for the internal chemical conversion~\cite{Siegel1995,roa2017catalyzed}. Importantly, our approach allows us to identify the governing timescales and to summarize the dynamical regimes in state diagrams with respect to physical and experimentally controllable parameters, such as microgel relaxation time, permeability, and the fuel concentration.

\begin{figure}[t]
  \includegraphics[width=\linewidth]{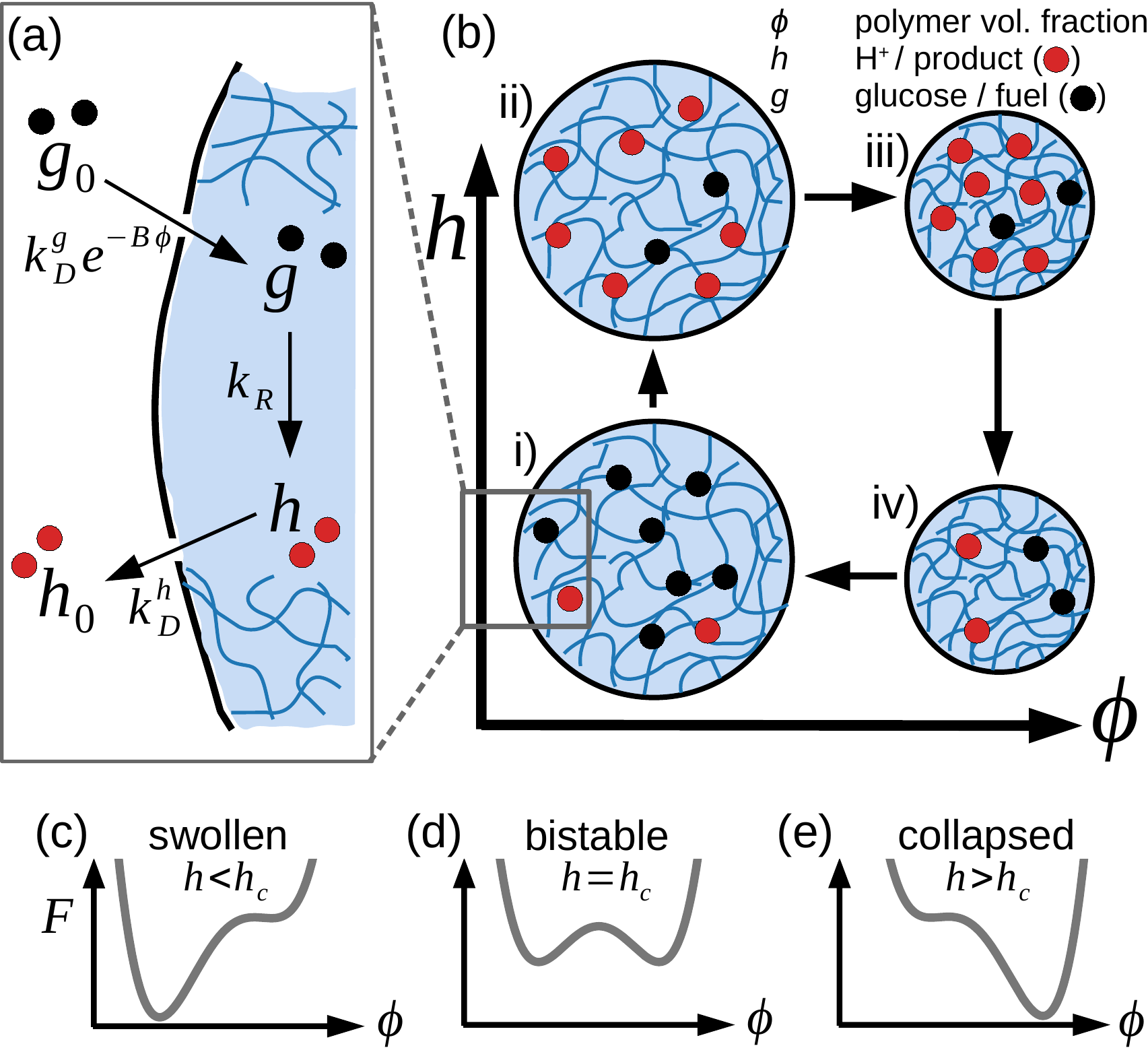}
  \caption{Model and feedback cycle. {\bfseries a)} Glucose (with outside concentration $g_0$) diffusively enters the gel according to a permeation rate $k_{\rm D}^g\exp(-B \phi)$ which depends on polymer density $\phi$ and produces protons $h$ with rate $k_R$ through catalysts (blueish background); protons $h$ diffuse out of the microgel with rate $k_{\rm D}^h$. {\bfseries b)} Sketch of the oscillation cycle i) to iv): The production of $h$ increases when the particle is swollen, i) to ii); $\phi$ decreases with large $h$ concentration, ii) to iii), inhibiting glucose uptake, iii) to iv), and thus yielding a negative feedback loop with \emph{activator} $h$ and \emph{inhibitor} $\phi$. Oscillations can only occur with \emph{spatial instabilities}, i.e., the volume phase transition of $\phi(h)$ must show hysteresis implying bistability in the free energy $F(\phi)$ (panels {\bfseries c)} to {\bfseries e)}) in the vicinity of the crossover concentration $h\C$. \label{fig:sketch}}
 \end{figure}

%
%
{\it Model and Methods.---} In our model, a reactant with a permeability-controlled, homogeneous inside concentration $g$ (e.g., glucose) converts within the microreactor by a catalytic process into inert non-relevant species and a product of concentration $h$ {(e.g., protons)} to which the gel volume is responsive, cf. Fig.~1(a). The gel volume is described by a time-dependent and intrinsically bistable polymer packing fraction $\phi$, assumed to be homogeneous in space. The feedback cycle i) to iv) of a single chemo-responsive microreactor is sketched and described in Fig.~\ref{fig:sketch}(b). Note that we use a sketch of spherical gels just for illustrative purposes; the model applies also to different finite microreactor geometries. Importantly, in the product-driven collapsed state, the large polymer density hinders transport of the fuel to the catalyst, thus imposing a time-delay negative feedback on the fueling process, where $h$ plays the role of the {\it activator} and $\phi$ the {\it inhibitor}. After the products have diffused out of the microgel the volume transition is reversed and can be re-fueled.
%

 To access the dynamics of such a chemomechanical feedback, we assume overdamped viscous dynamics of the gel volume fraction, via
 {$\tau_{\phi}   \dot\phi\propto -\beta {\partial F}/{\partial\phi}$},  
 where $F$ is the polymer's coarse-grained conformational free energy landscape, $\beta=(k\tb{B}T)^{-1}$ the inverse thermal energy, and $\tau_{\phi}$ the characteristic microgel swelling/shrinking relaxation time~\cite{Bell2021a,Richtering,Karg} in the linear response regime. To account for the typical hysteresis in the microgel switching response with respect to the stimulating concentration $h$~\cite{Baker1996}, we model $F(\phi)$ by a phenomenological Landau-like quartic form of double-well potential~\cite{Landau} (see Fig.~\ref{fig:sketch}(c)), resembling bistable Flory-Rehner free energies for a mean-field description of phase transitions~\cite{Flory,Boissonade2005,Yashin2012}. Further, we assume that the effects of products, $h$, on the volume transition can be described in first order by perturbations of $F(\phi)$ linear in $h-h_c$ and the `order parameter', $\phi$ (as the typical external field in the Landau theory), resulting in
 \begin{eqnarray}
  \tau_{\phi}\dot\phi&=&\left({\phi-\phi\C}\right) - a \left(\phi-\phi\C\right)^3+m(h-h\C), \label{eq:dotphi_0}
 \end{eqnarray}
where $h_c$ defines the critical concentration at which swollen and collapsed state are equally likely. The parameters $a$ and $\phi\C$ of the quartic potential as well as $m$ and $h\C$ can be obtained, in principle, by fitting Eq.~(\ref{eq:dotphi_0}) to experimentally accessible cosolute-driven hysteresis and transition curves (see the Supplemental Material~\cite{SI} for details). The linear perturbation term tilts the double-well potential [cf. Fig.~\ref{fig:sketch}(c)-(e)] and is well-established to describe the action of simple cosolutes on the coil-to-globule (or 2-state folding) transition of biomolecules within the popular linear `$m$-value' framework~\cite{Pace1975,Schellman1978,Heyda2014} and can also be justified for weakly charged electrostatic systems~\cite{heyda2014thermodynamic}.

The rate equations for the homogeneous glucose concentration inside the gel, $g(t)$, and the corresponding $\mathrm{H}^+$ concentration, $h(t)$, can be written as~\cite{SI}
\begin{subequations}\label{eq:hgkinetics}
\begin{eqnarray}
	\dot{g} &=& \kDg g_0 e^{-B\phi} - (\kDg + \kR) g,
	\label{eq:dotg_0}\\
	\dot{h} &=& \kR g - \kDh{(h-h_0)}.
	\label{eq:doth_0}
\end{eqnarray}
\end{subequations}
In Eq.~(\ref{eq:dotg_0}), $g$ is produced by diffusive influx with permeation {rate} $P=\kDg e^{-B\phi}$ ~\cite{Dhana,roa2017catalyzed}, where we approximate $\kDg$ with the diffusive rate in the limiting collapsed state~\cite{SI}. Note that in equilibrium ($\dot g=0$ and $\kR=0$), we obtain the correct partitioning $g/g_0={e^{-B\phi}}$. Early theory~\cite{Ogston,Gehrke}, supported by recent simulations~\cite{Q,Matej}, demonstrated the exponential dependence in $\phi$ with the `sieving' parameter, $B$, reflecting, e.g., the solute-to-polymer area ratio in the Ogston model~\cite{Ogston}. We restrict $B>0$, implying smaller partitioning for denser microgels. The volume dependence of the permeation rate, $P$, is key for the negative feedback in this system, including the hysteretic switch to block the fuel in the collapsed state~\cite{Zou1999,Leroux1999,Dhana,Bell2021a,Yao2017}. Moreover, in Eq.~(\ref{eq:doth_0}), $h$ is produced from $g$ with rate $k_R$ and diffusively leaves the gel by rate {$\kDh$} without any $\phi$-dependence. {We may further set $h_0=0$ without loss of generality as it invariantly shifts $h$ and $h\C$ while preserving the model dynamics.}

One can safely assume, that the small and free protons (H$^{+}$) diffuse significantly faster than the glucose can penetrate and react. Thus the dynamics of the inside proton concentration, $h(t)$, are limited by the glucose timescales, i.e., we obtain
$h = g\kR / \kDh$ from Eq.~(\ref{eq:doth_0}). Using the time derivative in Eq.~(\ref{eq:dotg_0}), $\dot{h} = \dot{g}~\kR / \kDh$,  we find
\begin{equation}
	\label{eq.eom_hplus}
	\dot{h} = \kDg \frac{\kR}{\kDh} g_0 e^{-B\phi} - (\kDg + \kR) h.
\end{equation}
We identify $(\kDg + \kR)^{-1} = \tau\tb{h}$ as the {\it effective} relaxation time of the proton concentration as well as the stationary distribution constant for the protons $K\tb{h}:= k/\kDh = {\kDg \kR }/[({\kDg + \kR}){\kDh}]$. The latter determines the steady-state H$^+$-concentration given a fixed $\phi$ and $g_0$ through $h=K\tb{h}g_0e^{-B\phi}$, while $k$ is the standard general rate constant in diffusion-influenced bimolecular reactions~\cite{roa2017catalyzed}. With that we arrive at the final kinetic equations for inhibitor and activator, $\phi$ and $h$,
%
%
%
\begin{subequations}
\begin{eqnarray}
 \tau_{\phi}\dot\phi&=&\left({\phi-\phi\C}\right) - a \left(\phi-\phi\C\right)^3+m(h-h\C),\label{eq:phi_final}\\
 \tau\tb{h}\dot{h}&=&K\tb{h}g_0e^{-B\phi}-h. \label{eq:h_final}
\end{eqnarray}
\label{eq:model_final}
\end{subequations}
We can now define the important timescale separation parameter $\epsilon={\tau_\phi}/{\tau\tb{h}}$. It quantifies the ratio between mechanical versus chemical relaxation times and is thus the key {physical parameter tuning} the dynamical behavior of the system.

%
%

\begin{figure}[t]
  \includegraphics[width=\linewidth]{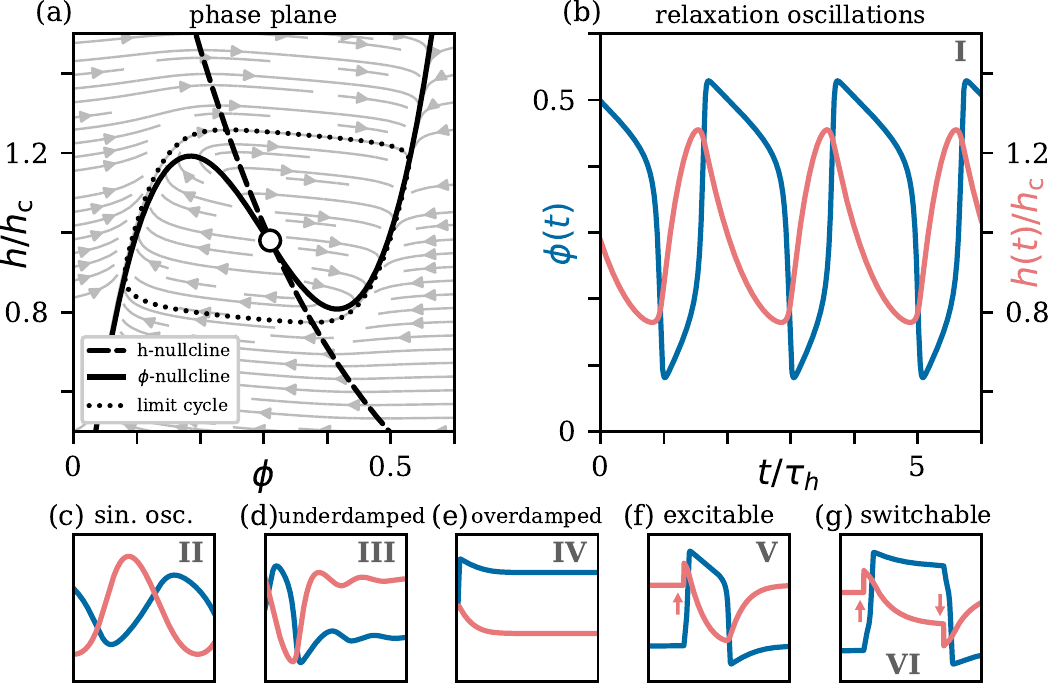}
 \caption{\label{fig:phase_plane} {\bfseries (a)}: Phase plane of Eq.~(\ref{eq:model_final}) depicting the $h$-nullcline (dashed), the $\phi$-nullcline (solid), the vector field (gray stream lines), and one example trajectory in the limit cycle (dotted) for relaxation oscillations with $K\tb{h} g_0/h\C =3$, $B=3.6$ and $\epsilon=0.05$. The corresponding time series for $h$ (red) and $\phi$ (blue) are presented in {\bfseries b)}. {\bfseries (c)}-{\bfseries (g)}: Time series of other behaviors for different sets of parameters. The parameter values are depicted in Fig.~\ref{fig:final_results}. The arrows in {\bfseries (f)} and {\bfseries (g)} indicate small external pertubations of $h$ into the shown direction. The roman numerals label the different behaviors.}
\end{figure}

{\it Results and Discussion.---} {Eq.~(\ref{eq:model_final}) can be studied geometrically in the planar phase space [see Fig.~\ref{fig:phase_plane}(a)] with straight-forward linear stability analysis~\cite{Strogatz,SI}. Remarkably, the system is conceptually identical with the FitzHugh-Nagumo (FHN) model, which is famous for modeling the rich neural excitation dynamics~\cite{Fitzhugh, Scott}. Compared to the FHN model, the polymer volume fraction, $\phi$, takes the role of the {\it nerve membrane potential} and the proton concentration, $h$, acts as the {\it recovery variable} (physiologically motivated by the voltage-gated, inactivating ion kinetics)~\cite{Izhikevich}. 
In fact, linearizing Eq.~(\ref{eq:h_final}) leads to a set of equations that is formally identical with the FHN model except for the opposite sign in Eq.~(\ref{eq:phi_final}) and an invariant shift of $\phi$ and $h$ by $\phi\C$ and $h\C$.}

{Because of the similarity of our system to the FHN model, one can rely on the extensive literature on {\it relaxation oscillators}~\cite{VanDerPol2009} and dynamics in neuroscience~\cite{Izhikevich,Keener2009} in general. Here, we reproduce the most probable dynamical regimes feasible for colloidal microreactors, and, most important, present phase diagrams with respect to experimentally meaningful system parameters.}

We fix the parameters $a=25$, $m={0.4/h\C}$, $\phi\C=0.3$, express all concentrations in terms of $h\C$ [Eq.~(\ref{eq:phi_final})], and demonstrate that the experimentally important parameters $K\tb{h}g_0$, $\epsilon=\tau_\phi/\tau\tb{h}$ and $B$ can be used to tune the system to exhibit the different dynamic behaviors presented as the typical nullcline analysis in Fig.~\ref{fig:phase_plane}. {In experiments $g_0$ is the fuel concentration, and $B$ the sieving parameter of the microgel~\cite{Ogston}. The timescale parameter $\epsilon$ could possibly be varied by changing the stiffness of the microgel, e.g., by tuning cross-link densities, and thereby its mechanical response time~\cite{Bell2021a,Richtering,Karg}.} 
 
\begin{figure}
 \includegraphics[width=0.9\linewidth]{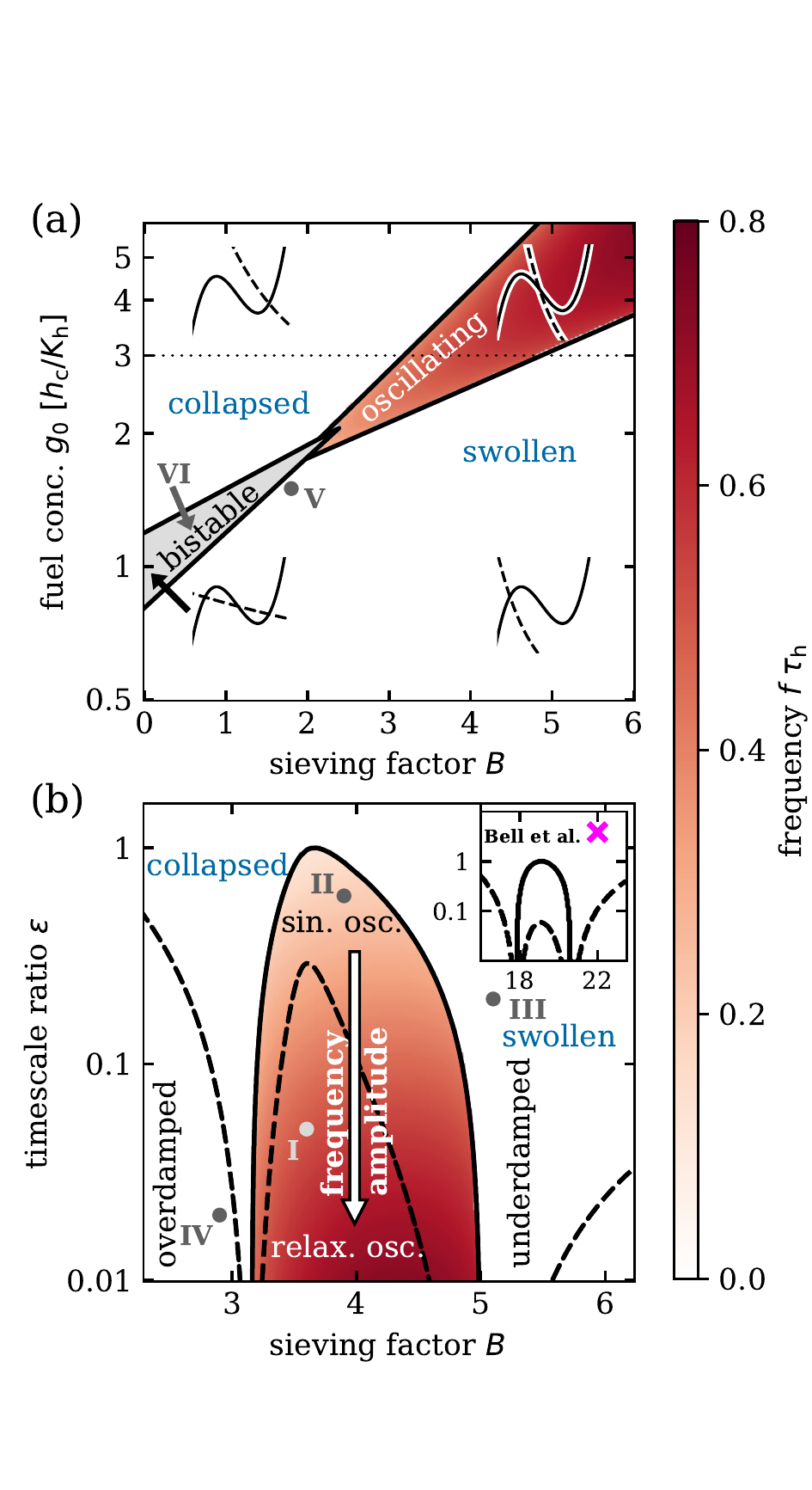}
 \caption{\label{fig:final_results}{\bfseries (a)}: Stability of Eq.~(\ref{eq:model_final}) in the $B$-$g_0$ plane with $\epsilon=0.05$. The system is either monostable (collapsed or swollen), bistable, or exhibits sustained oscillations (color-coded frequency obtained from numerical simulations). The insets sketch the nullclines and the fixed points in the $\phi$-$h$ phase plane. {Sustained oscillations can occur if the fixed point is on the central (decreasing) branch of the $\phi$-nullcline (cf. Fig.~\ref{fig:phase_plane}(a)), tuned by $B$.} Trajectories of an excitable (V) and a switchable state (VI) are shown in Fig.~\ref{fig:phase_plane}(f) and (g). {\bfseries (b)}: Details of the dynamic regimes in the $B$-$\epsilon$ plane with $g_0=3h\C/K\tb{h}$. Corresponding example trajectories {of states (I)-(IV)} are shown in Fig.~\ref{fig:phase_plane}(b)-(e).
 The time scale separation parameter, $\epsilon={\tau_\phi}/{\tau\tb{h}}$, needs to be at least smaller then unity for oscillatory behavior. In the monostable regime we find overdamped and underdamped oscillations. In the oscillatory regime, $\epsilon$ tunes the frequency (color-coded) and the amplitude. {The black lines originate from a stability analysis and separate the oscillatory and stable regime (solid), and different kinds of oscillatory/damped behavior (dashed); for details see~\cite{SI}. {The small inset displays the position of one paramater set (pink symbol: $B\approx 22, \epsilon\approx 4$), obtained from fitting experimental data \cite{Bell2021a,SI}, and the same bifurcation lines in the $\epsilon$-$B$-plane with $g_0\approx46h\C/K\tb{h}$ and $m\approx0.2/h\C$ for orientation. }  }
}
\end{figure}

The corresponding domains in parameter space are summarized in the state diagrams in Fig.~\ref{fig:final_results}. For relatively fast microgel relaxation, e.g., $\epsilon=0.05$ in Fig.~\ref{fig:final_results}(a), the microgel can be tuned from (monostable) collapsed to swollen. These two domains are separated by the bistable (small values of $K\tb{h}g_0$ and $B$) and the oscillating regime (large values) with the color-coded frequency. {We already conclude that high fuel sieving promotes oscillations, since the system can be driven into the oscillatory regime by increasing the fuel concentration. However, sufficient timescale separation is also required for sustained oscillations.}

The dashed line in Fig.~\ref{fig:final_results}(a) denotes $K\tb{h}g_0/h\C=3$, corresponding to panel (b), where more details with respect to $\epsilon$ are presented. The microgel must relax sufficiently fast, i.e., $\epsilon<1$, in order to observe oscillations at all. {In the oscillatory domain (with color-coded frequencies), we find small sinusoidal oscillations for $\tau_\phi\lesssim \tau\tb{h}$ with small frequencies ($f\propto1/\sqrt{\tau\tb{h}\tau_\phi}$), and a gradual cross-over to large high frequency ($f\propto1/\tau\tb{h}$) relaxation oscillations if the microgel responds instantaneously ($\tau_\phi\to0$) (see~\cite{SI} for details of the derivation of the scaling laws). This implies that the frequency is limited by the reaction and diffusion timescales ($\tau\tb{h}$) of the fuel and is further delayed by the microgel dynamics ($\tau_\phi$).}

{Finally, we demonstrate the explicit applicability of our model to  recent experimental works on spherical, catalytically active and pH-responsive gel particles using the glucose conversion~\cite{Bell2021a}. Since the cross-link ratio of these spheres is roughly 5 to 10\%, we assume a polymer volume fraction of $\phi\C\approx0.2$~\cite{Q,Sbeih2019}. The fit to our model~\cite{SI} predicts a  strong sieving ($B>20$) and further shows that $\tau\tb{h}\approx(18\pm 1)~\text{min}$, being significantly smaller than $\tau_\phi\approx(70\pm 10)~\text{min}$. According to our analysis, this leads to underdamped oscillations~\cite{SI}, as consistently observed in the experiments~\cite{Bell2021a}, in contrast to the desired sustained ones. {The position in parameter space is depicted in the inset of Fig.~3(b), indicating that slightly smaller sieving, or higher fueling, and faster microgel relaxation is necessary for sustained oscillations}. In order to stabilize the oscillations, we suggest to fine-tune the fuel concentration $g_0$, and, most important, reduce $\epsilon$, the ratio of the swelling kinetics vs. the glucose diffusion timescale $\kDg$, e.g., by manufacturing spheres of different sizes, use different copolymerization, or cross-link ratios. }
                                                                     
{\it Conclusion.---} Our work presents simple kinetic equations capturing the reaction-diffusion dynamics in a catalytic chemoresponsive microgel colloid, which is mathematically coherent with the FitzHugh-Nagumo model for prototypical excitable systems and also a realization of self-oscillating Ising model 'Gedankenexperiments'~\cite{DeMartino2019}. Compared to previous, more elaborate models of mechanochemical feedback, our model displays a convenient phase plane representation dismantling the system control parameters leading to {\it neural behavior}. The conceptual similarities of catalytically active, responsive microgels and established neural models marks a great potential for future intelligent soft matter devices. Further intriguing research is enabled towards smart and adaptive materials~\cite{Walther2020}, particularly, for the experimental control of the complex dynamics and communication of collectively coupled synergistic oscillators, such as in splay and chimera states,~\cite{chimera,PhysRevLett.130.107202,PhysRevLett.130.107201,Neural}, the inclusion of noise~\cite{Lindner,Lindner2000,Lindner1999,Levine2}, and the adaptation of our model to other feedback-controlled fueling sources, such as light~\cite{Ikkala}.

{The authors thank Ronald A. Siegel for helpful discussions and valuable advice.} This work was funded by the Deutsche Forschungsgemeinschaft (DFG) via the Research Unit FOR 5099 ``Reducing complexity of nonequilibrium systems''. The authors also acknowledge support by the state of Baden-Württemberg through bwHPC and the DFG through grant no INST 39/963-1 FUGG (bwForCluster NEMO) and under Germany's Excellence Strategy - EXC-2193/1 - 390951807 ('LivMatS').

	\bibliographystyle{apsrev4-2}
%

\end{document}